\documentclass[twocolumn]{aastex63}

\usepackage{lineno}
\usepackage[figuresright]{rotating}

\received{***, ****}
\revised{***, ****}
\accepted{***, ****}
\submitjournal{ApJ}

\shorttitle{Comparison of Helium Abundance}
\shortauthors{Song et al.}

\begin{document}
\title{Comparison of Helium Abundance between ICMEs and Solar Wind near 1 AU}

\correspondingauthor{Hongqiang Song}
\email{hqsong@sdu.edu.cn}

\author{Hongqiang Song}
\affiliation{Shandong Provincial Key Laboratory of Optical Astronomy and Solar-Terrestrial Environment, and Institute of Space Sciences, Shandong University, Weihai, Shandong 264209, China}
\affiliation{CAS Key Laboratory of Solar Activity, National Astronomical Observatories, Chinese Academy of Sciences, Beijing, 100101, China}



\author{Xin Cheng}
\affiliation{School of Astronomy and Space Science, Nanjing University, Nanjing, Jiangsu 210093, China}





\author{Leping Li}
\affiliation{CAS Key Laboratory of Solar Activity, National Astronomical Observatories, Chinese Academy of Sciences, Beijing, 100101, China}

\author{Jie Zhang}
\affiliation{Department of Physics and Astronomy, George Mason University, Fairfax, VA 22030, USA}

\author{Yao Chen}
\affiliation{Shandong Provincial Key Laboratory of Optical Astronomy and Solar-Terrestrial Environment, and Institute of Space Sciences, Shandong University, Weihai, Shandong 264209, China}







\begin{abstract}
The Helium abundance, defined as $A_{He}=n_{He}/n_{H}\times 100$, is $\sim$8.5 in the photosphere and seldom exceeds 5 in fast solar wind. Previous statistics have demonstrated that $A_{He}$ in slow solar wind correlates tightly with sunspot number. However, less attention is paid to the solar cycle dependence of $A_{He}$ within interplanetary coronal mass ejections (ICMEs) and comparing the $A_{He}$ characteristics of ICMEs and solar wind. In this paper we conduct a statistical comparison of Helium abundance between ICMEs and solar wind near 1 AU with observations of \textit{Advanced Composition Explorer} from 1998 to 2019, and find that the ICME $A_{He}$ also exhibits the obvious solar cycle dependence. Meanwhile, we find that the $A_{He}$ is obviously higher within ICMEs compared to solar wind, and the means within 37\% and 12\% of ICMEs exceed 5 and 8.5, respectively. It is interesting to answer where and how the high Helium abundance originates. Our statistics demonstrate that 21\% (3\%) of ICME (slow wind) $A_{He}$ data points exceed 8.5 around solar maximum, which decreases dramatically near minimum, while no such high $A_{He}$ values appear in the fast wind throughout the whole solar cycle. This indicates that the high $A_{He}$ (e.g., $>$8.5) emanates from active regions as more ICMEs and slow wind originates from active regions around maximum, and supports that both active regions and quiet-Sun regions are the sources of slow wind. We suggest that the high $A_{He}$ from active regions could be explained by means of the magnetic loop confinement model and/or photoionization effect.
\end{abstract}



\section{Introduction}
Coronal mass ejections (CMEs) are one of the most energetic eruptions in the solar system and can cause disastrous space weather effects \citep[e.g.,][]{gosling91,xumengjiao19}. It is widely accepted that CMEs result from eruption of magnetic flux ropes \citep[e.g.,][]{chenpengfei11}, which can form prior to \citep[e.g.,][]{chengxin11,patsourakos13} and during \citep[e.g.,][]{song14a,jiangchaowei21} eruptions. The composition (including both the charge states and elemental abundances of heavy ions) provides an important avenue to analyze their eruption process \citep{lepri04, lynch11,gruesbeck11,gruesbeck12,song15b,song16,song17a,song20b,guchaoran20,song21b}, and supports that more CMEs come from active regions around solar maximum, whose heavy ions possess higher charge states and enriched relative abundances \citep{song21a}.

Solar wind is a stream of charged particles released from the corona, and can be divided into fast and slow streams taking speed as criterion. It is generally accepted that fast wind originates from coronal holes, and slow wind can emanate from active regions and/or quiet-Sun regions \citep[e.g.,][]{zhaoliang17a,fuhui18}. Two classes of models have been proposed to explain the origin of solar wind, including the wave-turbulence-driven (WTD) models, which suggest the solar wind being released along open magnetic field lines directly \citep{hollweg86,cranmer07,verdini09}, and the reconnection loop opening (RLO) models, which propose the stream escaping through magnetic reconnections between open and closed magnetic field lines \citep{fisk99,fisk03,woo04} and work for solar wind originating from closed-field region. These models can be examined through analyzing the solar wind composition.

The Helium abundance, defined as $A_{He}=n_{He}/n_{H}\times 100$, can be employed to analyze the source regions and release processes of both CMEs and solar wind. As Hydrogen and Helium are the first and second most abundant elements within both interplanetary coronal mass ejections (ICMEs) and solar wind \citep{aellig01,kasper07,kasper12,mcintosh11,alterman19,alterman21}, respectively, they can be measured in situ with relatively higher cadence and accuracy compared to the other heavier ions.

$A_{He}$ is $\sim$8.5 in the photosphere \citep{asplund09} and seldom exceeds 5 in fast solar wind \citep{kasper07,kasper12} partly because the first ionization potential (FIP) of Helium (24.6 eV) is higher than that of Hydrogen (13.6 eV). This makes the neutral Helium atoms harder to be ionized through thermal ionization and flow out than the Hydrogen in the lower solar atmosphere, resulting in the $A_{He}$ depletion in the corona and solar wind \citep{laming04,laming15}. Earlier statistics also demonstrated that the $A_{He}$ in slow wind correlates tightly with sunspot number (SSN) \citep{aellig01,kasper07,kasper12,mcintosh11,alterman19}, and can be used to herald the solar cycle onset \citep{alterman21}. Compared to the solar wind, the statistical studies reveal that the $A_{He}$ is obviously higher within ICMEs \citep{owens18,huangjin20}.

It is interesting to explore where and how the high $A_{He}$ of ICMEs originates. \cite{fuhui20} suggested that the chromospheric evaporation during solar eruptions can supply plasma with high $A_{He}$ to CMEs through a case study, while this scenario can not consistently explain that ICMEs usually exhibit the obvious FIP effect, \textit{i.e.}, possessing higher Fe/O, Mg/O, and Si/O than the photosphere and chromosphere \citep{zurbuchen16,owens18,huangjin20}. Therefore, it might not work well for most events. Previous studies showed that the $A_{He}$ is a linear function of solar wind speed around solar minimum, and the correlation between $A_{He}$ and SSN is strongest at low speed, while the positive correlation decreases quickly with increasing speed \citep{kasper07,kasper12}, which can be explained qualitatively by means of the coronal loop confinement model \citep{fisk03,woo04,endeve05,kasper07}. This model connects the confinement time of plasma in a loop to its physical size, and these factors in turn influence the temperature profile, gravitational settling, and plasma composition \citep{kasper07}. Besides, \cite{song21a} reported that Ne/O exhibits opposite solar cycle dependence within ICMEs and solar wind, and they suggested that the photoionization might be responsible for the higher Ne/O ratios within ICMEs around solar maximum.

We anticipate that a statistical comparison of $A_{He}$ between ICMEs and solar wind could shed more light on whether the loop confinement model and photoionization can be employed to answer the origin of high $A_{He}$ within ICMEs. This is the major motivation for us to conduct this study. The paper is organized as follows. In Section 2, we introduce the data and ICME catalogs, and the statistical results and comparisons are displayed in Section 3, which is followed by the discussion and summary as the final section.

\section{Data and ICME Catalogs}
The Helium abundance ($A_{He}$) data in this study are provided by the Solar Wind Electron Proton Alpha Monitor \citep[SWEPAM;][]{mccomas98} experiment aboard \textit{Advanced Composition Explorer (ACE)}, which orbits around the L1 point since launched in 1997 \citep{stone98}. The SWEPAM observations are made with independent electron and ion instruments through electrostatic analyzers. With the fan-shaped fields of view, SWEPAM can sweep out all pertinent look directions as \textit{ACE} spins, and provide 64-s resolution of the full electron and ion distribution functions \citep{mccomas98}. All of these data can be downloaded at the ACE science center\footnote{http://www.srl.caltech.edu/ACE/ASC/level2/index.html}. The yearly SSNs are obtained at the Solar Influence Data Center of the Royal Observatory of Belgium\footnote{http://www.sidc.be/silso/home}.

ICMEs can be identified with different physical features, such as enhanced magnetic field intensity, smoothly changing field direction, low proton temperature, low plasma $\beta$ and so on \citep[e.g.,][]{wucc11}. Researchers have identified ICMEs through in-situ data near 1 AU and offered several complete and reliable catalogs using measurements of the \textit{ACE} \citep[RC catalog]{richardson10}, \textit{WIND} \citep{chiyutian16,nieves18}, as well as the \textit{Solar Terrestrial Relations Observatory} \citep{jianlan18}. Every catalog lists the ejecta boundaries of each ICME, which are adopted to analyze the Helium abundance within ICMEs. In this paper, the RC catalog\footnote{http://www.srl.caltech.edu/ACE/ASC/DATA/level3/icmetable2.html} is used as the $A_{He}$ is provided by the SWEPAM aboard the \textit{ACE}.

\section{Statistical Results and Comparisons}
RC catalog lists 487 ICMEs totally from 1996 to 2019, and the yearly ICME numbers are displayed with white histograms as shown in Figure 1. To demonstrate the solar cycle dependence of ICME numbers, the yearly SSNs are also plotted here with the black line. The Spearman rank cross-correlation coefficient ($\rho$) between yearly ICME numbers and SSNs is calculated with the IDL code $r\_correlate.pro$, and $\rho >$ 0.6 implies the meaningful cross correlation \citep[e.g.,][]{alterman19}. The $\rho$ between yearly numbers of ICMEs and sunspots is 0.85, which is also shown in Figure 1 and illustrates a strong correlation between them.

The SWEPAM began to provide $A_{He}$ from 1998 February onward, and data gaps appear occasionally during the normal observations. Thus the $A_{He}$ is available just for 420 ICMEs that are depicted with light grey in Figure 1, including 155 (37\%) and 50 (12\%) ICMEs with their average $A_{He}$ exceeding 5 and 8.5, which are displayed with the dark grey and black histograms, respectively. Figure 1 shows that ICMEs with high $A_{He}$ values mainly appear during solar maximum and descending phases, similar to the relative abundances of the other heavier ions \citep{song21a}.

To analyze the solar cycle dependence of ICME $A_{He}$, we first calculate the average value of $A_{He}$ within each ICME with the 64-s resolution SWEPAM data. The average duration of ICMEs is $\sim$20 hours \citep[e.g.,][]{wucc11}, which means one ICME contains $\sim$1125 data points on average. Then the yearly mean, median, and standard deviation of all the ICMEs in each year are calculated. In the meantime, we also calculate the yearly values for slow and fast solar wind. The interaction between fast streams, emanating from coronal holes, and slow streams can produce the Co-rotating Interaction Regions (CIRs) in interplanetary space \citep{jianlan06b}, where the solar wind speeds can be changed obviously. For example, the original slow ($<$400 km s$^{-1}$) and fast ($>$600 km s$^{-1}$) wind are accelerated and decelerated to $\sim$500 km s$^{-1}$ (see Figure 2 in \cite{jianlan06b} for an example). If simply set 400 or 500 km s$^{-1}$ as the critical value to differentiate fast and slow wind, the original slow/fast wind might be counted as fast/slow wind given the existence of CIRs. Therefore, we set our slow and fast wind threshold at $<$400 km s$^{-1}$ and $>$600 km s$^{-1}$, i.e., excluding the wind between 400 and 600 km s$^{-1}$, which could guarantee more that pure slow or fast streams are investigated without contamination from each other. Table 1 lists all of the calculated results for ICMEs, slow wind, and fast wind.

\begin{center}
\begin{sidewaystable}[thp]
\centering
\caption{The yearly sunspot number (SSN), average value (mean), median (median) and standard deviation (stddev) of $A_{He}$ within ICMEs, slow solar wind (SSW) and fast solar wind (FSW) in each year from 1998 to 2019.}
\tabcolsep=2pt
\begin{tabular}{cccccccccccccccccccccccc}
\tableline
\tableline
     &Year    & 1998 & 1999 & 2000 & 2001 & 2002 & 2003 & 2004 & 2005 & 2006 & 2007 & 2008 & 2009 & 2010 & 2011 & 2012 & 2013 & 2014 & 2015 & 2016 & 2017 & 2018 & 2019\\
     &SSN     & 88.3 & 136.3& 173.9& 170.4& 163.6& 99.3 & 65.3 & 45.8 & 24.7 & 12.6 & 4.2  & 4.8  & 24.9 & 80.8 & 84.5 & 94.0 & 113.3& 69.8 & 39.8 & 21.7 & 7.0  & 3.6 \\
\tableline
ICME &mean    &3.918 &3.968 &5.682 &5.294 &6.073 &6.194 &5.016 &5.848 &2.902 &2.856 &1.622 &1.266 &2.986 &3.532 &4.831 &5.082 &5.540 &4.577 &3.084 &3.477 &1.919 &2.045\\
     &median  &3.476 &3.660 &4.860 &4.919 &5.479 &5.474 &4.252 &5.875 &1.786 &2.856 &1.622 &0.731 &2.056 &2.832 &4.023 &4.176 &4.585 &3.686 &2.266 &3.165 &1.655 &0.931\\
     &stddev  &2.313 &2.458 &3.130 &2.608 &3.487 &3.399 &2.952 &3.605 &2.039 &0.463 &1.206 &1.293 &2.179 &2.645 &3.475 &3.686 &3.269 &2.434 &2.088 &1.742 &1.183 &1.788\\
SSW  &mean    &2.651 &3.252 &4.092 &3.705 &4.576 &4.776 &3.486 &2.884 &2.171 &1.950 &1.345 &1.382 &2.110 &3.107 &3.358 &3.007 &3.844 &3.550 &2.383 &2.382 &1.759 &1.528\\
     &median  &2.280 &2.910 &3.840 &3.340 &4.310 &4.230 &2.990 &2.610 &1.920 &1.620 &0.960 &1.090 &1.540 &2.950 &3.100 &2.810 &3.770 &3.490 &2.150 &2.190 &1.510 &1.070\\
     &stddev  &1.713 &1.991 &2.022 &2.183 &2.404 &2.763 &2.207 &1.662 &1.357 &1.413 &1.227 &1.080 &1.952 &1.768 &1.922 &1.673 &1.592 &1.608 &1.326 &1.300 &1.208 &1.270\\
FSW  &mean    &2.284 &2.425 &2.885 &2.098 &2.362 &2.171 &2.567 &2.514 &2.917 &3.118 &2.372 &3.101 &4.036 &4.099 &4.084 &3.227 &3.201 &3.510 &2.489 &3.219 &2.836 &2.717\\
     &median  &2.240 &2.310 &2.610 &1.450 &2.290 &1.940 &2.390 &2.330 &2.870 &3.030 &2.330 &3.070 &4.000 &4.010 &4.000 &3.090 &3.060 &3.610 &2.360 &3.150 &2.760 &2.635\\
     &stddev  &1.509 &1.292 &1.956 &2.105 &1.029 &1.130 &1.573 &1.114 &1.341 &1.116 &1.063 &0.576 &1.766 &1.505 &0.850 &1.540 &1.184 &1.351 &1.019 &1.009 &1.298 &0.904\\
\tableline
\tableline
\end{tabular}
\end{sidewaystable}
\end{center}

Figure 2 displays the yearly means of $A_{He}$ within ICMEs (pink), slow wind (blue), and fast wind (red), along with the SSNs (black, also listed in Table 1) together, and the corresponding Spearman rank coefficient of correlation between $A_{He}$ and SSN are 0.85, 0.90, and -0.22, which demonstrate quantitatively that the Helium abundance within ICMEs also possess the solar cycle dependence while fast wind does not. As only a few ICMEs were detected in each year around solar minimum, we do not divide the yearly ICMEs into magnetic clouds and non-cloud events, or fast and slow ones. \cite{owens18} reported that magnetic clouds have higher $A_{He}$ than non-cloud ICMEs. In the meantime, the median and standard deviation are not plotted in Figure 2 for clarity, which could be found in Table 1.

Here the strong correlation between $A_{He}$ and SSN in the slow wind is consistent with previous statistical results with WIND measurements \citep{kasper07,kasper12,alterman19,alterman21}. One different point is that $A_{He}$ of solar wind from \textit{WIND} has comparable values during maximums of solar cycles 23 and 24, although the SSN amplitude of cycle 24 is less compared to previous cycle, see Figure 1 in \cite{alterman19}. While the \textit{ACE} measurements demonstrate that the $A_{He}$ values of both slow wind and ICMEs around cycle 24 maximum (2012--2015) decreased slightly compared to cycle 23 (2000--2003), consistent with the expectation of SSN variation \citep{alterman21}.

The quantitative comparison between the two solar maximums shows that the average $A_{He}$ of slow wind (ICMEs) decreases by 16.3\% (11.7\%), i.e. from 4.16 (6.16) to 3.48 (5.44), and the t-test \citep[see e.g.,][]{song21a} results illustrate that the variation is significant at the 99\% confidence level. The discrepancy of $A_{He}$ variation trends between \textit{WIND} \citep{alterman19} and \textit{ACE} should mainly result from different selections of velocity intervals. \cite{alterman19} presented the $A_{He}$ of solar wind with velocity between 312 and 574 km s$^{-1}$, while our velocity range is less than 400 km s$^{-1}$ as mentioned. We can get the same result that the Helium abundances of solar wind are comparable during the last two solar maximums if choosing the same velocity interval with \cite{alterman19} using \textit{ACE} data (not shown).

Figure 3 presents a series of normalized distributions of $A_{He}$ data points with 64-s resolution for ICMEs (top), slow wind (middle), and fast wind (bottom). Figure 2 has shown that the $A_{He}$ of ICMEs and slow wind can change significantly from solar maximum to minimum, while we find that no obvious differences exist between the overall distributions of $A_{He}$ for solar cycles 23 and 24. Therefore, the two solar cycles (1998--2019) are displayed together in left panels, and the two solar maximum (2000--2003 \& 2012--2015) and minimum (2006--2009 \& 2016--2019) phases are presented in the middle and right panels, respectively. The red vertical lines denote the photospheric value of $A_{He}$ (i.e., 8.5). The quantitative analysis shows that 21\% of data points within ICMEs possess high $A_{He}$ value ($>8.5$) around solar maximum, which decreases dramatically to 2.3\% around minimum. For the solar wind, only slow wind around maximum has 2.7\% of data points that exceed 8.5, and in other cases the high $A_{He}$ ($>8.5$) data points could be negligible.

The mean and median of the $A_{He}$ distributions are also displayed in each panel of Figure 3. It shows that ICMEs possess the highest $A_{He}$ compared to slow and fast wind, consistent with Figure 2. Focus on the solar wind alone, the $A_{He}$ is more enriched in fast wind compared to slow wind around solar minimum, while slow wind possesses higher $A_{He}$ than fast wind around maximum. The $A_{He}$ of fast wind does not exhibit the solar cycle dependence, while slow wind possesses obvious higher $A_{He}$ around maximum compared to minimum. This implies that slow winds originating from active region and quiet-Sun region have different properties, as slow winds mostly come from quiet-Sun region around solar minimum, and more fractions of slow wind from active region around maximum.

\section{Discussion and Summary}
Our study demonstrates that $A_{He}$ in both ICMEs and slow wind exhibits the positive correlation with SSNs, which indicates that the high $A_{He}$ (e.g., $>$8.5) emanates from active regions as more ICMEs and slow wind originates from active regions around solar maximum. In the meantime, no high $A_{He}$ ($>8.5$) data points existing in fast wind throughout a solar cycle implies that coronal holes do not emanate plasmas with enriched Helium. This enlighten us to infer how the high $A_{He}$ originates through comparing the characteristics of active regions and coronal hole.

It is generally accepted that the dominated magnetic field lines in active regions and coronal holes are closed and open, respectively. Some studies displayed that the high $A_{He}$ could be explained through the coronal loop confinement model \citep{fisk03,woo04,endeve05,kasper07}. This model suggested that the properties of coronal loop plasmas can vary over their confinement duration before the plasmas are released through magnetic reconnection between loop and open field lines \citep{fisk03}. As mentioned, the confinement time of plasmas in the coronal loop is correlated with its physical size, and these factors in turn influence the temperature profile, gravitational settling, and composition of the plasma. For example, simulations \citep{killie05} displayed that the $A_{He}$ in closed field regions can increase obviously in several days due to the thermal force, which is caused by the energy dependence of the Coulomb cross section. The thermal force can be large in the transition region and seeks to push Helium ions toward corona, causing the buildup of Helium in coronal loop \citep{killie05}. These high $A_{He}$ plasma confined in the coronal loop can be involved into CMEs through reconnection occurring in the current sheet beneath CMEs during solar eruptions \citep{linjun00,linjun04,song16}, corresponding to the enriched Helium within ICMEs. Also, they can be released into slow wind by reconnection between closed loop and open field lines.

The other potential mechanism for high $A_{He}$ within ICMEs is related to photoionization, similar to explaining high Ne/O ratios within flare regions and ICMEs \citep{feldman05,zurbuchen16}. According to the FIP effect \citep{laming04,laming15}, the Ne/O should decrease in the corona and ICMEs compared to the photosphere as the FIP of Ne (21.6 eV) is higher than O (13.6 eV). However, measurements showed the opposite case within flare regions and ICMEs \citep{feldman05,song21a}. \cite{shemi91} has suggested that pre-flare soft X-ray can penetrate through the chromosphere and create a slab-like region with photoionization ratios at the chromosphere base. As the photoionization cross section ratio of Ne and O is 9:4 \citep{yeh85}, the ionization ratio of Ne is higher than that of O. This leads to plasma being transported into corona with higher Ne/O, corresponding to the enriched Ne/O in flare regions \citep{schmelz93} and ICMEs \citep{song21a}. Meanwhile, more CMEs are accompanied by energetic flares during solar maximum, which indicates that the photoionization might play a more important role around maximum compared to minimum, further enhancing the solar cycle dependence of ICME Ne/O. Similarly, the photoionization cross section of He is larger than that of H \citep{yeh85}, thus it is straightforward to speculate that the photoionization leads to the enhanced $A_{He}$ within ICMEs and flare regions \citep{feldman05}. Simulations are needed to examine this conjecture.

In this paper we conducted a statistical comparison of Helium abundance between ICMEs, slow solar wind ($<$400 km s$^{-1}$) and fast solar wind ($>$600 km s$^{-1}$) near 1 AU with \textit{ACE} observations from 1998 to 2019, covering solar cycles 23 and 24. The statistics demonstrated that the $A_{He}$ of ICMEs and slow wind exhibits the obvious solar cycle dependence, and the $A_{He}$ within ICMEs is highest compared to slow and fast winds. The averages within 37\% and 12\% of ICMEs exceed 5 and 8.5, respectively. About 21\% (3\%) of $A_{He}$ data points of ICMEs (slow wind) exceed 8.5 around solar maximum, which decreases dramatically around minimum, while no such high $A_{He}$ values appear in the fast wind throughout the solar cycle. Focus on the solar wind alone, the $A_{He}$ is more enriched in fast wind compared to slow wind around solar minimum, while slow wind possesses higher $A_{He}$ than fast wind around maximum. This indicates that high $A_{He}$ originates from active regions. Two possible mechanisms are discussed to explain the high $A_{He}$ within both slow wind and ICMEs around solar maximum.

\acknowledgments We thank the anonymous referee for the comments and suggestions that helped to improve the original manuscript. We acknowledge the use of ICME catalog provided by Richardson \& Cane. All the SWICS data are downloaded from the \textit{ACE} science center. Hongqiang Song thanks Drs. Bo Li, Liang Zhao, and Liang Guo (a statistician at Shandong University) for their useful discussions. This work is supported by the NSFC grants U2031109, 11790303 (11790300), and 12073042. Hongqiang Song is also supported by the CAS grants XDA-17040507 and the open research program of the CAS Key Laboratory of Solar Activity KLSA202107.





\begin{figure*}[htb!]
\epsscale{0.9} \plotone{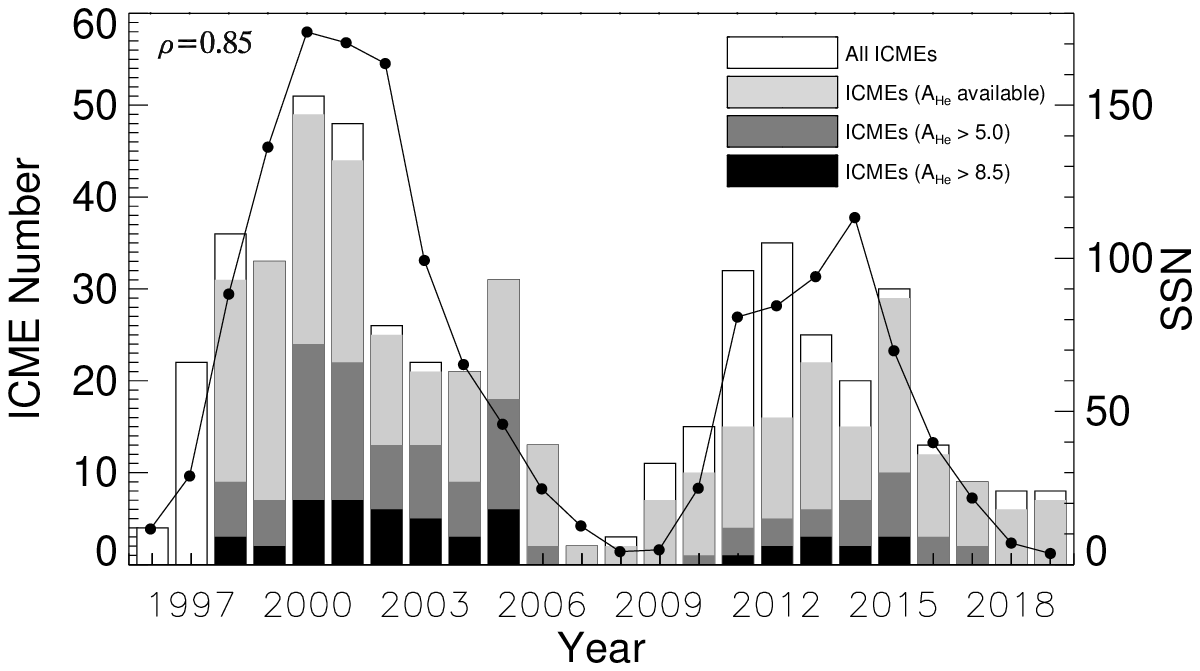} \caption{Yearly numbers of ICMEs at L1 point and sunspots from 1996 to 2019. The white histograms represent the ICME numbers with the light grey depicting those with available $A_{He}$. The dark grey and black regions indicate the ICME numbers with $A_{He}$ exceeding 5 and 8.5, respectively. The black line displays the yearly sunspot numbers. The Spearman rank cross-correlation coefficient ($\rho$) between yearly numbers of ICMEs and sunspots is also presented. \label{Figure 1}}
\end{figure*}

\begin{figure*}[htb!]
\epsscale{0.9} \plotone{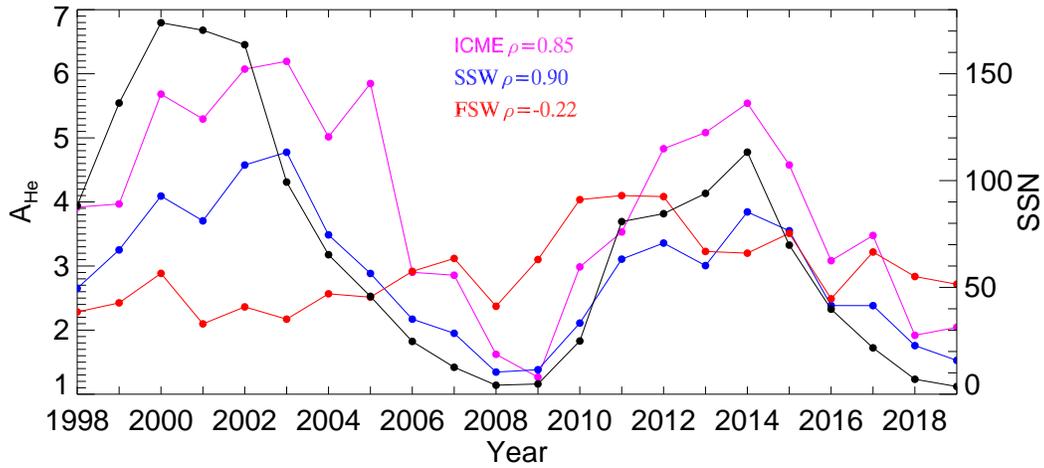} \caption{The solar cycle dependence of $A_{He}$ within ICMEs (pink), slow solar wind (SSW, blue) and fast solar wind (FSW, red). The Spearman rank cross-correlation coefficients ($\rho$) between yearly $A_{He}$ and SSNs (depicted with black line) are also presented. \label{Figure 2}}
\end{figure*}

\begin{figure*}[htb!]
\epsscale{1.0} \plotone{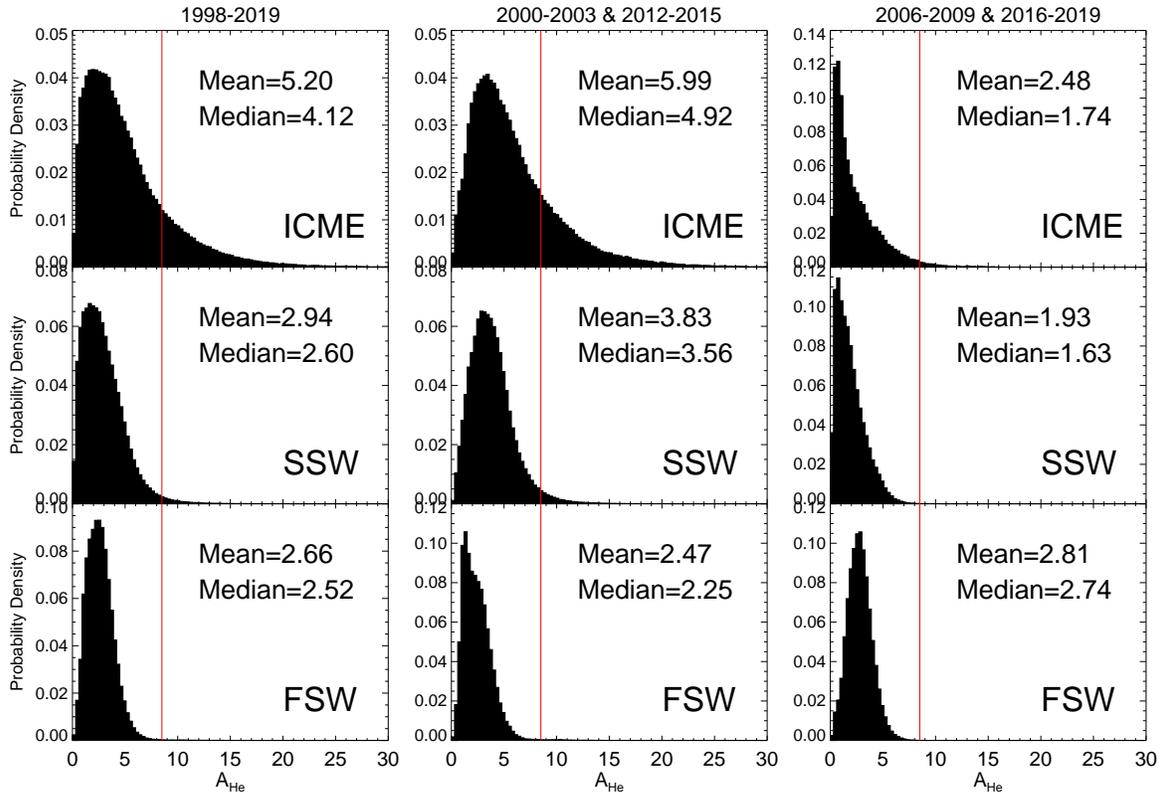} \caption{The normalized distributions of $A_{He}$ data points with 64-s resolution within ICMEs (top), slow solar wind (SSW, middle), and fast solar wind (FSW, bottom). The two solar cycles (1998--2019) are displayed together in left panels, and the two solar maximum (2000--2003 \& 2012--2015) and minimum (2006--2009 \& 2016--2019) phases are presented in the middle and right panels, respectively. The corresponding mean and median are given in each panel, and the vertical red lines depict the $A_{He}$ value in the photosphere. \label{Figure 3}}
\end{figure*}

\end{document}